\documentclass[a4paper]{spie}  

\newcommand{\lb}{\emph{LiteBIRD}}

\usepackage{amsmath,amsfonts,amssymb}
\usepackage{graphicx}
\usepackage{hyperref}
\hypersetup{colorlinks,allcolors=black}
\usepackage{amsmath} 
\usepackage[capitalise]{cleveref}
\usepackage{amsmath,amsfonts,amssymb}

\title{Systematic effects induced by half-wave plate differential optical load and TES nonlinearity for LiteBIRD}

\author[a]{Silvia Micheli}
\author[b,c]{Tijmen de Haan}
\author[c]{Tommaso Ghigna}
\author[a]{Alessandro Novelli}
\author[a]{Francesco Piacentini}
\author[a]{Giampaolo Pisano}
\author[a]{Fabio Columbro}
\author[a]{Alessandro Coppolecchia}
\author[a]{Giuseppe D'Alessandro}
\author[a]{Paolo de Bernardis}
\author[a]{Luca Lamagna}
\author[a]{Elisabetta Marchitelli}
\author[a]{Silvia Masi}
\author[a]{Andrea Occhiuzzi}
\author[a]{Alessandro Paiella}

\affil[a]{Dipartimento di Fisica, Università La Sapienza, P.le A. Moro 2, I-00185 Roma, Italy and INFN Sezione di Roma, P.le A. Moro 2, I-00185 Roma, Italy}

\affil[b]{Institute of Particle and Nuclear Studies (IPNS), High Energy Accelerator Research Organization (KEK), Tsukuba, Ibaraki 305-0801, Japan}
\affil[c]{International Center for Quantum-field Measurement Systems for Studies of the Universe and Particles (QUP-WPI), High Energy Accelerator Research Organization (KEK), Tsukuba, Ibaraki 305-0801, Japan }

\authorinfo{E-mail: silvia.micheli@uniroma1.it}

\begin{document} 
\maketitle

\begin{abstract}
\lb, a forthcoming satellite mission, aims to measure the polarization of the Cosmic Microwave Background (CMB) across the entire sky. The experiment will employ three telescopes, Transition-Edge Sensor (TES) bolometers and rotating Half-Wave Plates (HWPs) at cryogenic temperatures to ensure high sensitivity and systematic effects mitigation. This study is focused on the Mid- and High-Frequency Telescopes (MHFT), which will use rotating metal mesh HWPs. We investigate how power variations due to HWP differential emissivity and transmittance combine with TES nonlinear responsivity, resulting in an effective instrumental polarization. We present the results of simulations for the current HWP design, modeling the TES deviation from linearity as a second-order response. We quantify the level of acceptable residual nonlinearity assuming the mission requirement on the tensor-to-scalar ratio, $\delta r < 0.001$. Moreover, we provide an accuracy requirement on the measurement of TES responsivity nonlinearity level for MHFT channels. Lastly, we present possible mitigation methods that will be developed in future studies.

\end{abstract}


\section{Introduction}
\label{sec:intro}  

Precision cosmology has experienced an accelerating development in the last decades, and observations of the Cosmic Microwave Background (CMB) temperature and polarization have led to unprecedented accuracy in the estimation of cosmological parameters\cite{2020planck}. However, accurate measurements of the CMB polarization can still answer many questions about the early Universe. Measurements of the faint primordial \textit{B}-mode polarization would indeed provide information about cosmic inflation, parameterized by the tensor-to-scalar ratio, $r$\cite{2016Kamionkowski}. For this reason, several ground-based\cite{2019ade, Abazajian_2022} and balloon-borne\cite{2018gualtieri,2014dober,2021addamo} experiments are targeting the \textit{B}-mode signal. The \lb\ space mission has been selected to observe the signature of inflation from space. \lb\ is a satellite mission led by JAXA, planned to be launched in the Japanese fiscal year of 2032\cite{ptep}. It will target the \textit{B}-mode signal, observing the CMB across fifteen frequency channels, from 34 to 448 GHz, scanning the whole sky with its three telescopes, (Low-, Mid-, and High- Frequency Telescopes), aiming to reach a sensitivity of $\delta r < 10^{-3}$. This type of measurement faces a number of challenges, such as handling the Galactic foregrounds at large angular scales and the lensed signal over the recombination bump. Moreover, several systematic effects can contaminate polarization measurements. A key element in the experimental setup of \lb\ is a continuously rotating half-wave plate (HWP). A sapphire HWP will be employed in the polarization modulation unit (PMU) of LFT, while metal-mesh HWPs will be used as the first optical element for MFT and HFT. A HWP is a birefringent medium geometrically tuned to generate a  phase delay between light polarized along its ordinary and extraordinary axes. A continuously rotating HWP at a rate $f_{\mathrm{HWP}}$ modulates the incoming polarized signal at $4f_{\mathrm{HWP}}$, suppressing the 1/$f$ noise and allowing single detector measurements of Stokes parameters. However, using a rotating HWP can introduce systematic effects that need to be carefully addressed. These effects can arise from imperfections in the HWP material, thermal instability of the HWP, nonideal behavior of the rotation mechanism, or variations in the incident angle of the CMB radiation\cite{giardiello2022,sugiyama2020}. \\ \\
\lb\  focal planes will be equipped with polarization sensitive Transition-Edge Sensor (TES) bolometers. We simulate the impact of TES nonlinearity on the timestreams using the MNTES tool \cite{2024tdh}, showing the effect of saturation. We show how nonidealities of the HWP, including the HWP synchronous signal (HWPSS) coming from the HWP differential optical properties, couple to TES nonlinearity. In particular, we investigate the effective instrumental polarization resulting from TES nonlinearity in the presence of a large HWPSS. We include these effects in time-ordered data (TOD) simulations using the \url{litebird_sim} (publicly available at \url{https://github.com/litebird/litebird_sim}) simulation framework, focusing on the impact on MHFT channels. We discuss the implications of the coupled effects on the estimation of the tensor-to-scalar ratio, setting a requirement on the level of knowledge of TES responsivity nonlinearity level.

\section{HWP differential optical load}

\subsection{MHFT HWP design}
\begin{figure} [ht]
   \begin{center}
   \begin{tabular}{c} 
   \includegraphics[width=0.5\textwidth]{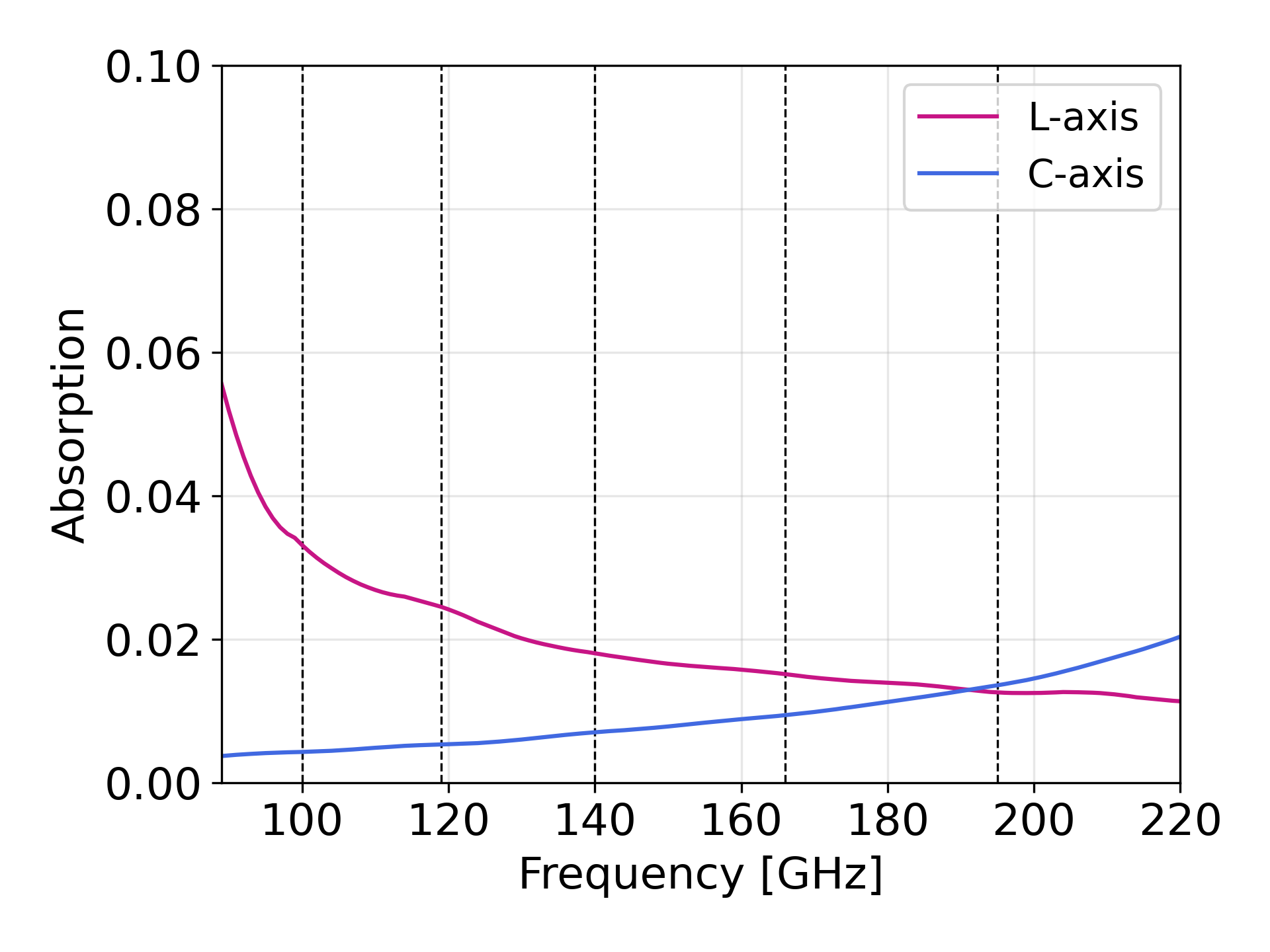}
   \includegraphics[width=0.5\textwidth]{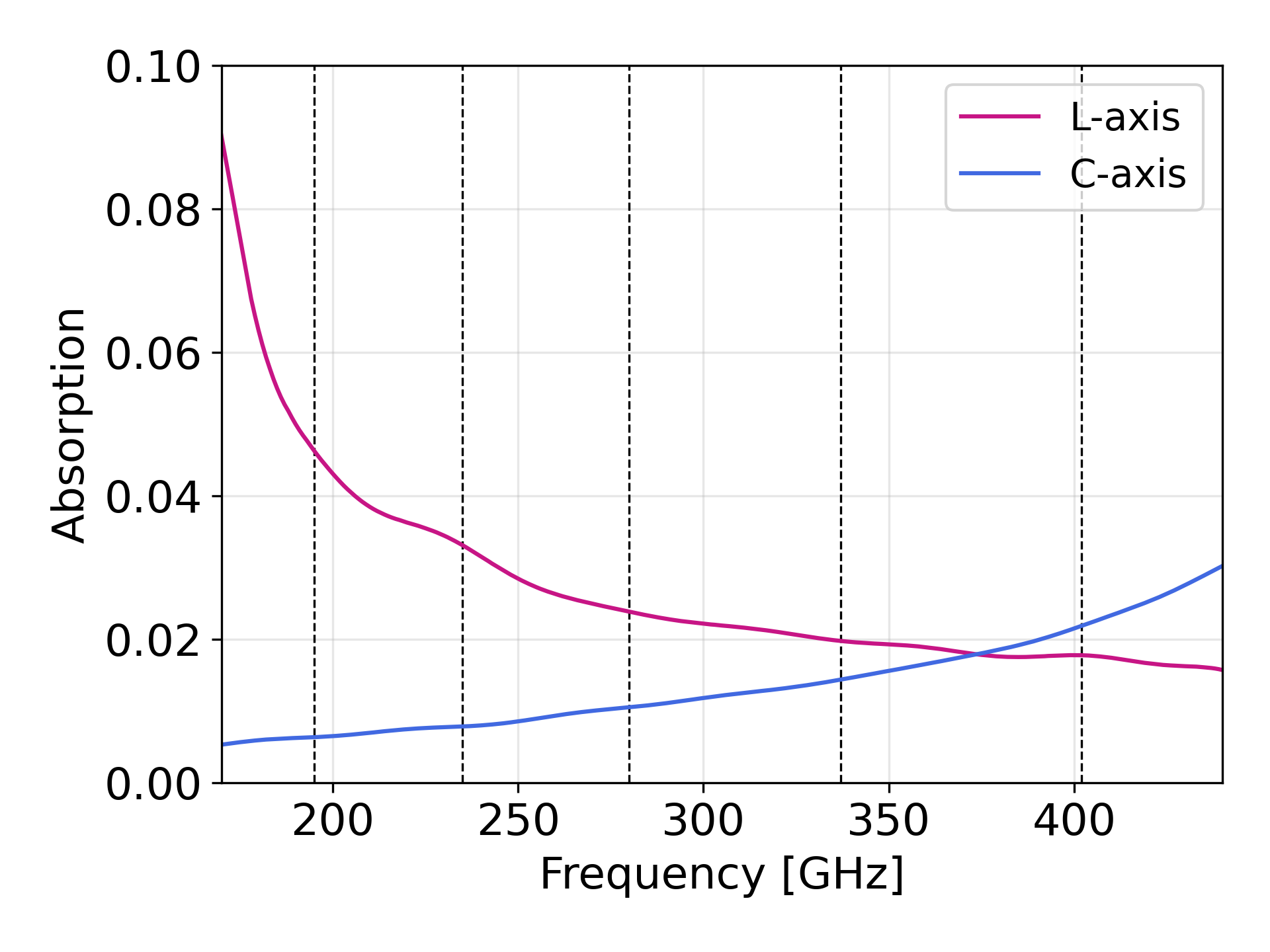}
   \end{tabular}
    \begin{tabular}{c} 
   \includegraphics[width=0.5\textwidth]{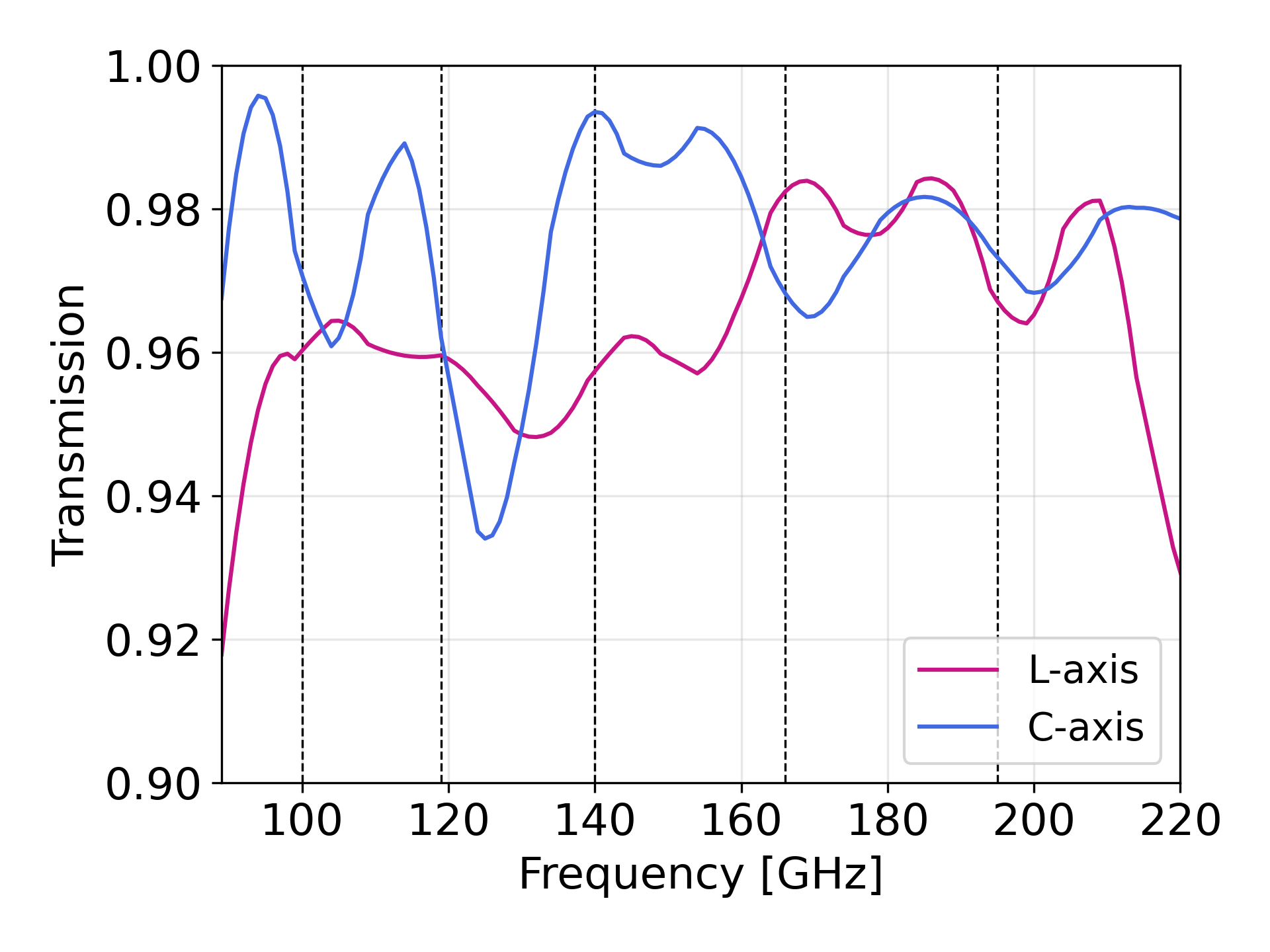}
   \includegraphics[width=0.5\textwidth]{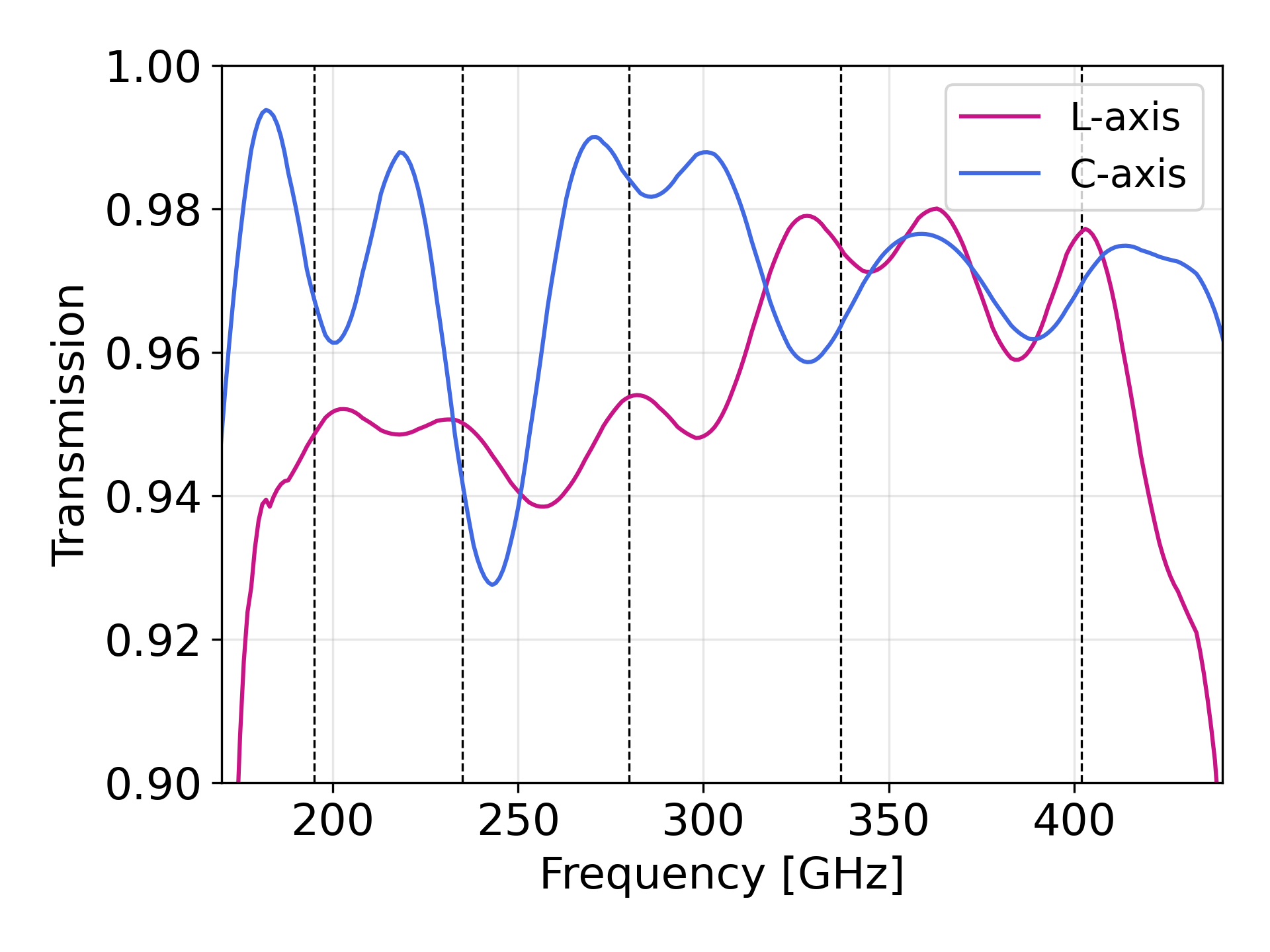}
   \end{tabular}
   \end{center}
   \caption[mhft] 
   {\label{fig:mhft_values}MFT (\emph{left}) and HFT (\emph{right}) HWPs absorptions (\emph{upper}) and transmissions (\emph{lower}) as a function of frequency, for the capacitive (blue) and inductive (purple) axes. The central frequency for each channel is shown by the dashed lines.}
\end{figure}

Using a rotating HWP effectively reduces numerous systematic effects. When the HWP modulates the sky signal, the CMB polarization appears in the sidebands of the $4f_{\mathrm{HWP}}$ modulation signal, well above the 1/$f$ noise component, which is then naturally suppressed. Moreover, fast rotation of the HWP ($f_{\mathrm{HWP}}$ = 0.65 and 1.02 Hz, for MFT and HFT, respectively) also allows quasi-instantaneous single-detector measurements of the $I$, $Q$, and $U$ Stokes parameters. This eliminates the necessity to compensate for differences among detectors, such as gain and beam mismatches. However, strong parasitic signals, which peak at the harmonics of the HWP rotation frequency, $f_{\mathrm{HWP}}$, can result from HWP imperfections. We refer to these signals as HWP synchronous signals (HWPSS). A detailed study of HWPSS correction methods was performed for previous experiments such as EBEX\cite{Araujo2017}, NIKA\cite{2017ritacco} and POLARBEAR\cite{takakura2017}, presenting a number of sources for the HWPSS as well. In this work, we focus on the $2f_{\mathrm{HWP}}$ polarized signal arising from the HWP differential transmission and emission, resulting from different optical properties along its ordinary and extraordinary axes. \\ The PMUs for MFT and HFT will employ spinning metal-mesh HWPs, that emulate the behavior of birefringent materials, located at the 20-K temperature stage. These devices are made of orthogonally oriented stacks of capacitive and inductive metal grids embedded in polypropylene. The specific combination of stacks provides a $180^{\circ}$ phase-shift around the operational frequency range\cite{2015pisano}. The expected performance of the MFT and HFT HWPs\cite{2022pisano,2023columbro} are reported in \cref{fig:mhft_values}. Here, we show how the absorption and transmission along the two axes of the HWP change in frequency. 


\subsection{Effect of differential absorption and emission}\label{sec:hwp_diff}
\begin{figure} [ht]
   \begin{center}
   \begin{tabular}{c} 
   \includegraphics[width=0.5\textwidth]{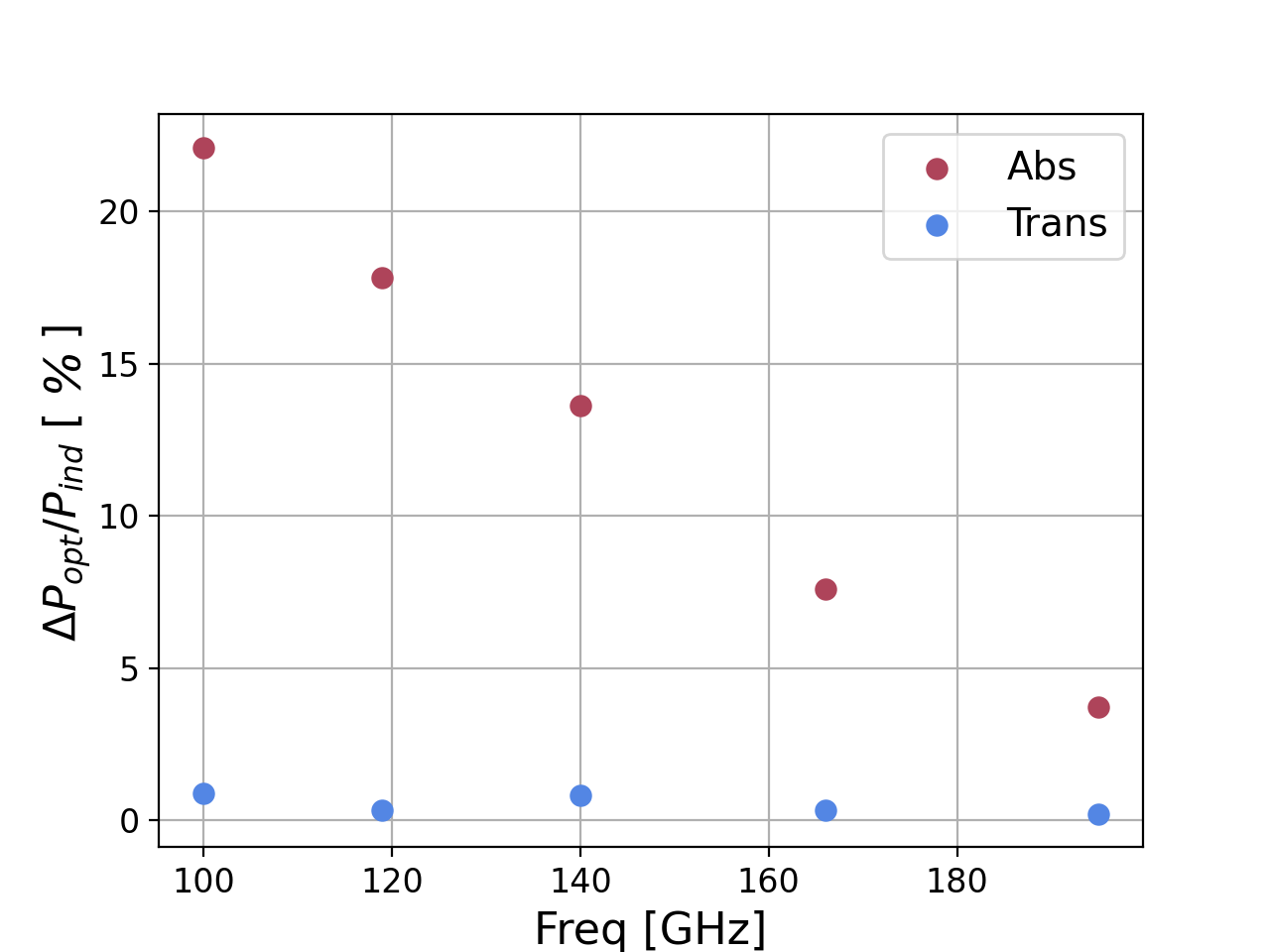}
   \includegraphics[width=0.5\textwidth]{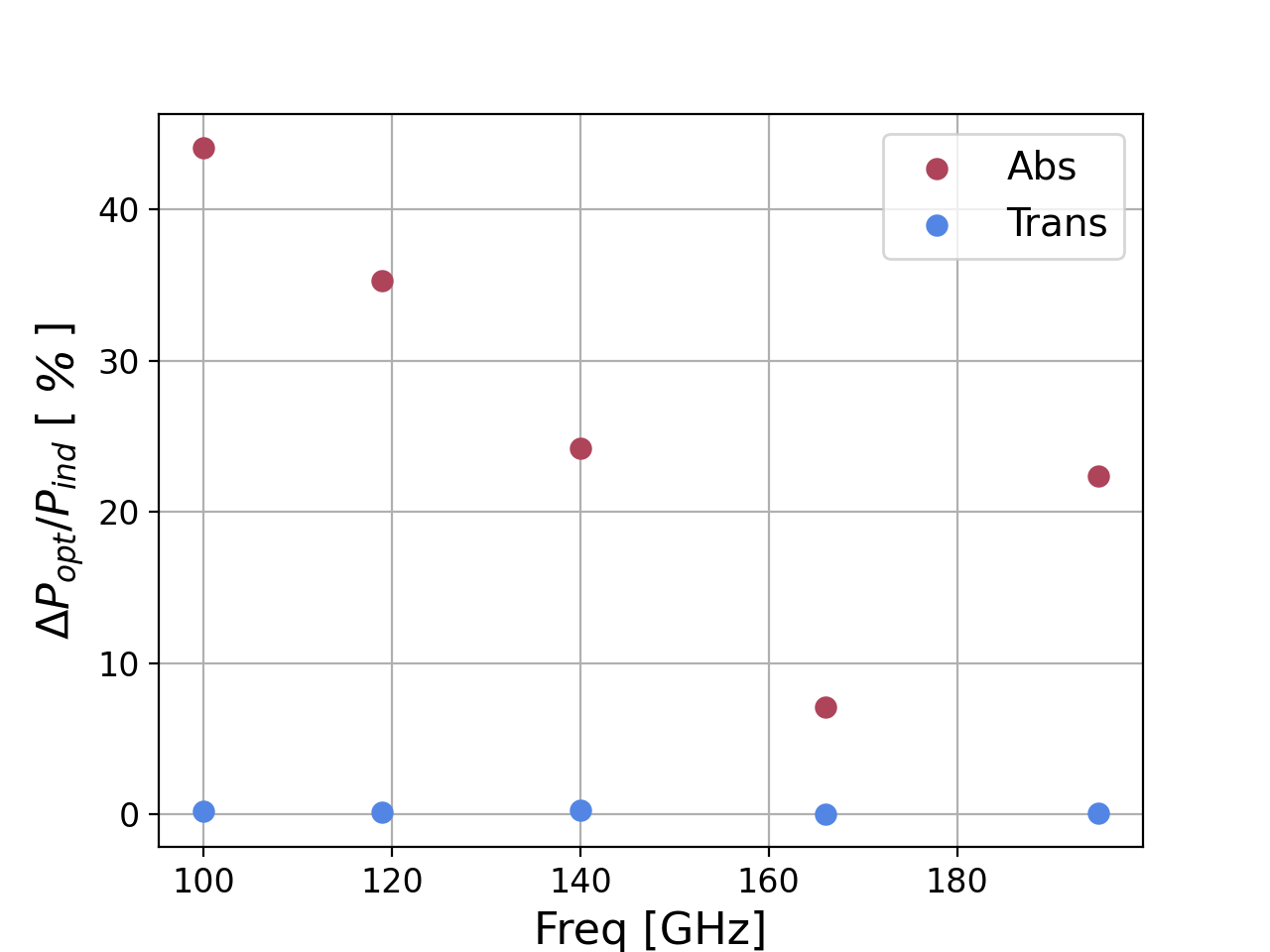}
   \end{tabular}
   \end{center}
   \caption[mhft] 
   {\label{fig:mhft_diff}: MFT \emph{(left)} and HFT \emph{(right)} differential absorption (red) and transmission (blue) contributions to power variation $\Delta P_{\mathrm{opt}}$ with respect to the power along the inductive axis, $P_{ind}$, which is, in the majority of the cases, set as the nominal value.}
\end{figure}

\begin{figure} [ht]
   \begin{center}
   \begin{tabular}{c} 
   \includegraphics[width=0.5\textwidth]{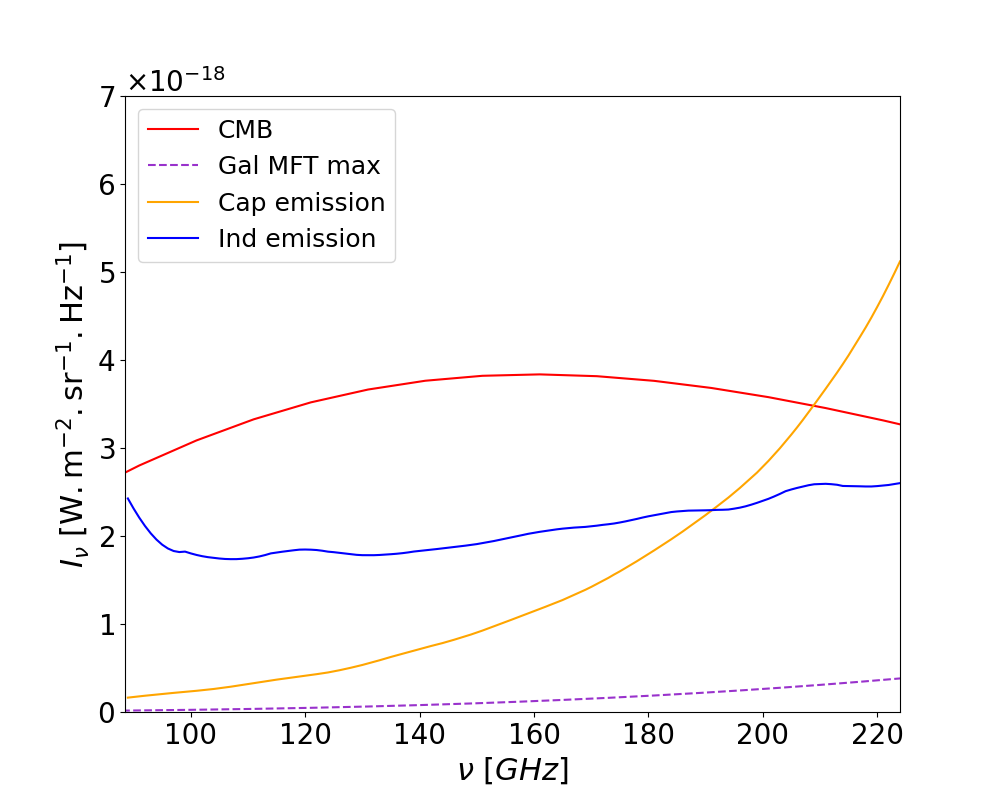}
   \includegraphics[width=0.5\textwidth]{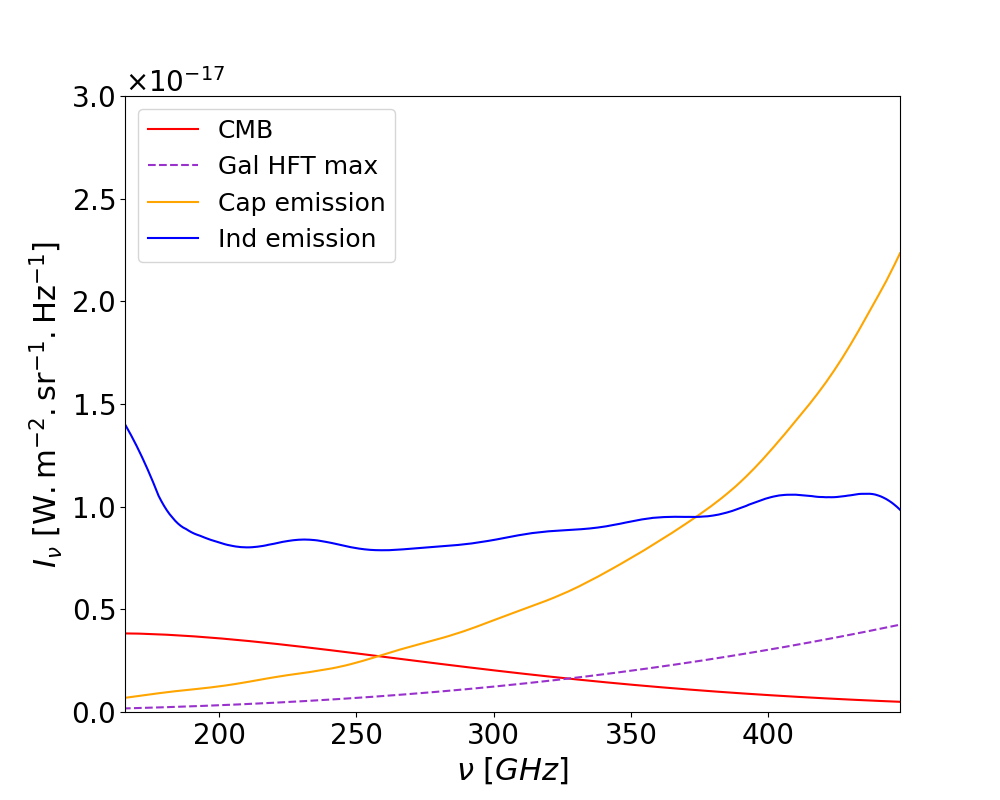}
   \end{tabular}
   \end{center}
   \caption[emisgal] 
   {\label{fig:emisgal}Representation of different signal contribution for MFT (\emph{left}) and HFT (\emph{right}) frequency ranges. The blue and orange lines show the gray body emission of the 20-K HWP along the inductive and capacitive axis, respectively. The purple dashed line represents the maximum spectral density for the thermal dust emission from the Galaxy and the red solid line shows the CMB temperature spectrum.}
\end{figure}

We simulate the effect of different values of transmittance and emissivity of the HWP on the nominal optical power value using a code based on the sensitivity model for \lb\cite{hasebe2022}, modeling the HWPs performance as shown in \cref{fig:mhft_values}. Thus, we obtain the optical power variation expected in each channel, with respect to the nominal values.\\As can be observed in \cref{fig:mhft_diff}, the major contribution for MHFT bands comes from differential emission (or absorption), whereas a small percentage of signal comes from differential transmission. Similar studies were conducted for LFT as well\cite{2022ghigna,2023patanchon}, focusing on the role of differential transmission, which is instead dominant at low frequencies. In this work, we will use ideal HWP performances for LFT. For high frequency bands, it is also worth noting that the optical power coming from HWP differential emission is often even higher than the maximum Galactic emission, in the respective bands. A qualitative impression of this is shown in \cref{fig:emisgal}. 
As the HWP rotates by an angle $\chi=2\pi f_{\mathrm{HWP}} t$, the HWPSS can be written as the sum of the HWP harmonics, as\cite{salatino2018}:
\begin{equation}
    A(\chi) = \sum_{n=1}^{\infty} (A_n + a_n^{opt}I)\cos(n\chi + \phi_n)
\end{equation}
Here, the amplitude of each harmonic $n$ is expressed by the sum of $A_n$, which is constant in time during the scan, and a component that varies with the incoming intensity, with coefficient $a_n^{opt}$. As shown above, the impact of differential emission exceeds the one of differential transmission by several orders of magnitude in MHFT channels, so this study will exclusively address the former. Hence, we will express the modulation from the HWPSS as:
\begin{equation}\label{eq:hwpss}
    d_{\text{HWPSS}} = A_2 \cos (2\chi)
\end{equation}
where the amplitude of the signal is given by $A_2 = \Delta P_{\mathrm{HWP}}$. As a consequence, the signal hitting each detector will be:
\begin{equation}
    d(t) = I + \mathrm{Re}[\epsilon_{\mathrm{pol}}e^{4i\chi}(Q+iU)]+A_2 \cos(2\chi) + N
\end{equation}
where $I$, $Q$ and $U$ are the Stokes components of the incoming radiation, $\epsilon_{\mathrm{pol}}$ is the polarization efficiency and \textit{N} is the noise contribution.
\section{TES nonlinearity}
\subsection{TES response model}\label{sec:tes_nl}
\lb\ will employ nearly 5000 TES bolometers in order to reach the required statistical sensitivity. A TES operates by measuring current at fixed voltage. The current is a measurement proxy for incident optical power via the change in resistance of the superconducting film held in its superconducting-to-normal transition. This resistance change steeply varies with the temperature of the film, making the TES an extremely sensitive thermometer. The measured current depends on the incident power nonlinearly. The first derivative of the relation is defined as the responsivity:
\begin{equation}
    S \equiv \frac{\partial I}{\partial P_{\mathrm{opt}}}
\end{equation}
Responsivity inherently varies with optical power, $P_{\mathrm{opt}}$. The nonlinear relation between current and power is determined in detail in the companion paper de Haan et al. (2024, in preparation \cite{2024tdh}). In this work, we assume a second order approximation of the MNTES curve to model a residual nonlinearity after correction. In particular, we want to assess the acceptable level of residual nonlinearity, i.e. how well we should be able to correct for it, in order not to affect the estimation of $r$. \\ \\ The analysis presented in this work employs full TOD simulations, which are produced in units of K. In order to convert the optical power from the HWPSS, we calculate a conversion factor for each channel, according to the following steps:
\begin{itemize}
    \item compute $P_{\mathrm{opt}}$ on the detectors for $T_{\mathrm{CMB}}$ (nominal optical power)
    \item compute $P_{\mathrm{opt}}$ on the detectors for $T^*_{\mathrm{CMB}} = T_{\mathrm{CMB}} + \delta T$, with $\delta T  = 1$ mK
    \item compute the optical power variation associated with the temperature variation $\delta T$
        \begin{equation}
            \delta P_{\mathrm{opt}} = P_{\mathrm{opt}}(T^*_{\mathrm{CMB}}) - P_{\mathrm{opt}}(T_{\mathrm{CMB}})
        \end{equation}
    \item calculate the conversion factors from power to temperature units:
        \begin{equation}
            C [K/W] = \frac{\delta T}{\delta P_{\mathrm{opt}}}
        \end{equation}
\end{itemize}
We use these coefficients to convert the HWPSS from W to K units and include it in the \url{litebird_sim} framework. Note that this procedure is not related to calibration techniques, rather it just provides quick unit conversion factors. In order to inject the nonlinearity in the TODs, we follow the first order approximation described in the POLARBEAR analysis\cite{takakura2017}. Thus, we introduce the nonlinearity as a second order correction to the responsivity (gain), modifying the signal as:
\begin{equation}\label{eq:nl}
    d_{NL}(t) = [1+g_1d(t)]d(t)
\end{equation}
where $d(t)$ and $d_{NL}(t)$ are in K units, and the second order gain correction $g_1$ is in K$^{-1}$ units. Typically, $g_1$ is a negative value\cite{didier2019}. We will refer to $g_1$ as the \textit{nonlinearity parameter} in this work. The time constant of the detector can vary as well, causing a different nonlinear effect\cite{ghigna2023}, which will be discussed in a future paper.

\subsection{Effect of nonlinearity}
\begin{figure} [ht]
   \begin{center}
   \includegraphics[width=0.5\textwidth]{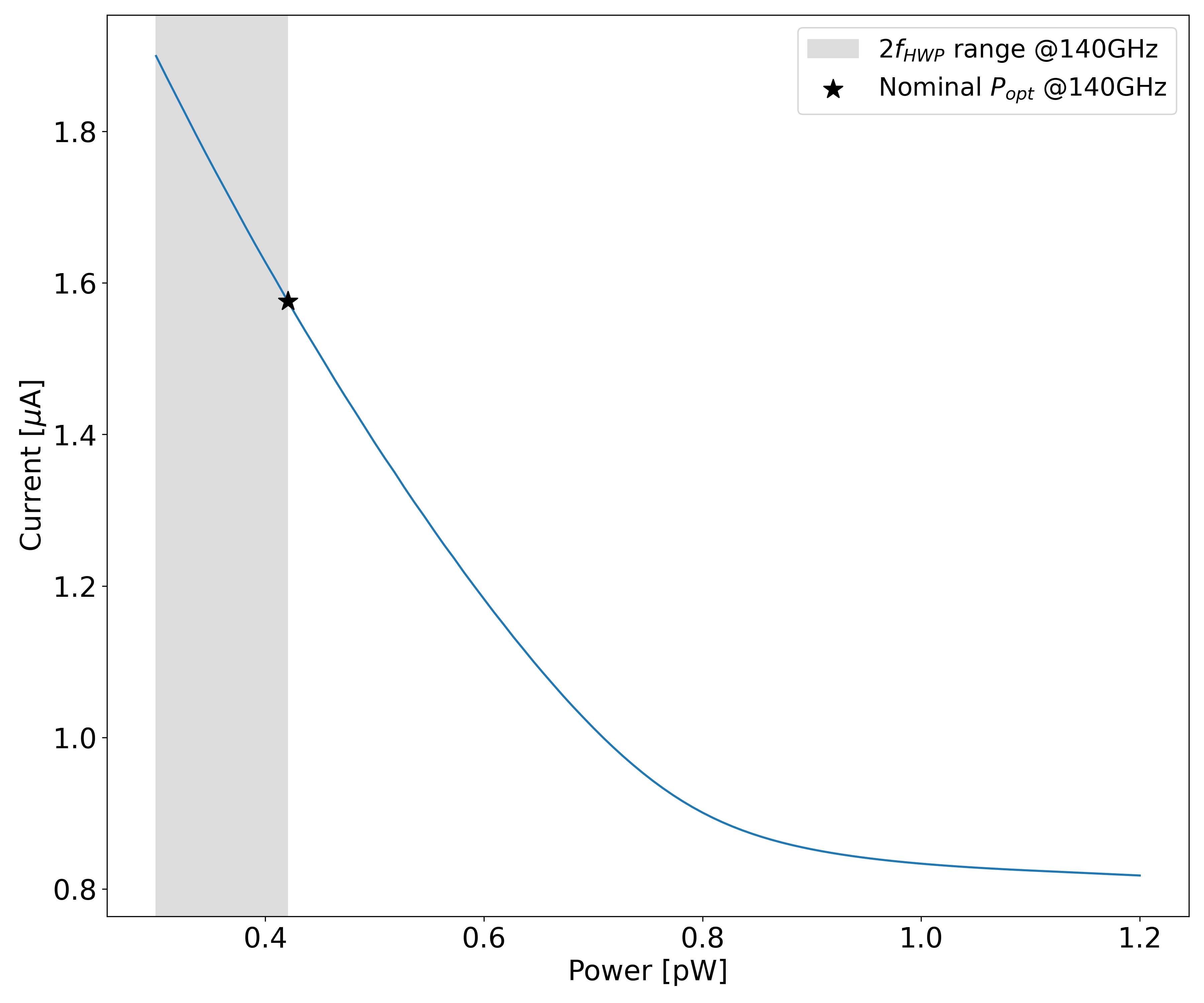}
   \end{center}
   \caption[tescurve] 
   {\label{fig:tes_curve}TES response curve reconstructed with the MNTES tool. The detector saturates as the optical power increases. We show the nominal optical power value for the M1-140 frequency band and its variation range in the presence of HWPSS.}
\end{figure}

\begin{figure} [ht]
   \begin{center}
   \includegraphics[width=0.5\textwidth]{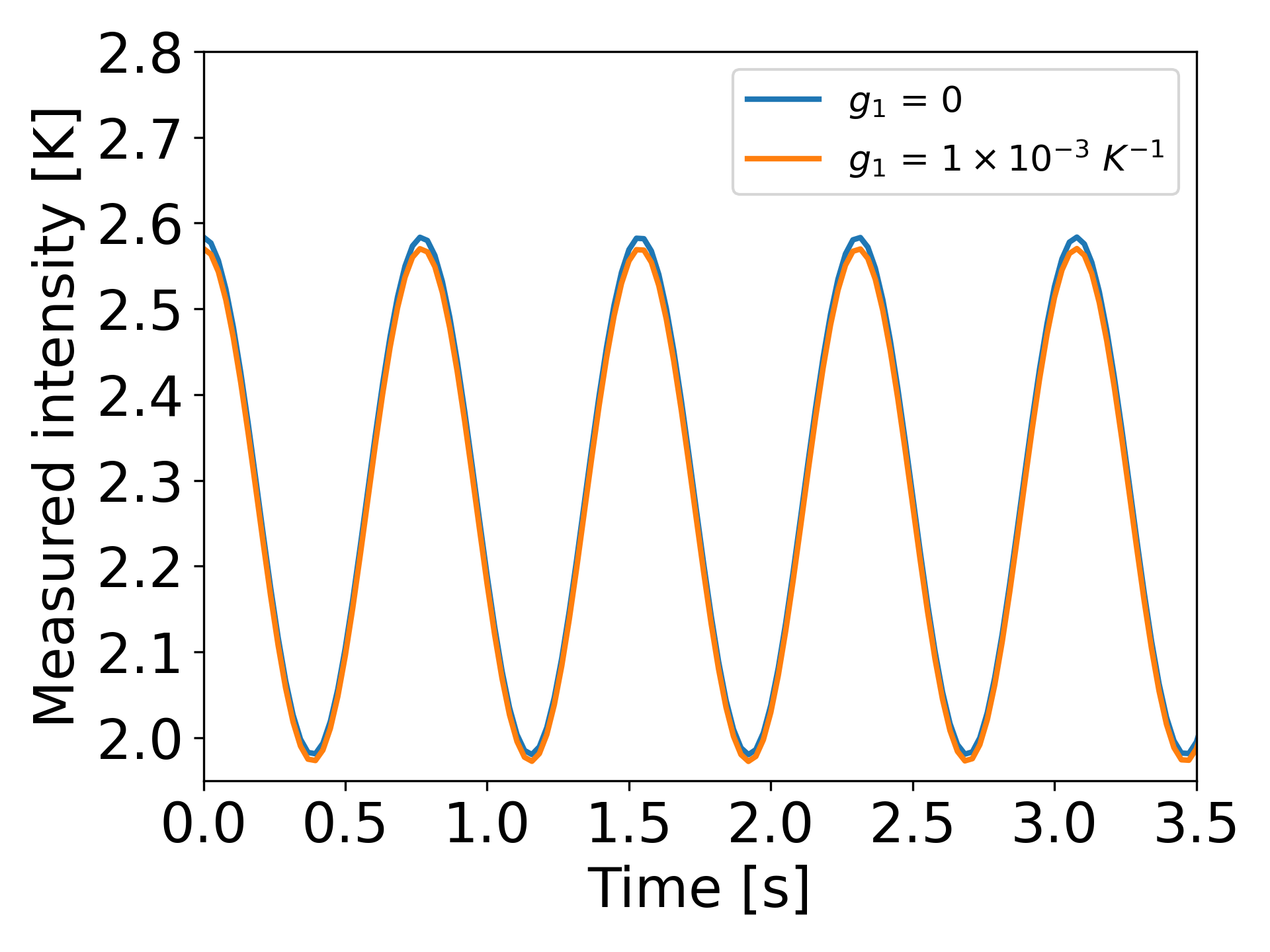}
   \end{center}
   \caption[todcomp] 
   {\label{fig:tod_comp}Comparison of a chunk of simulated TOD, after applying TES nonlinear response (orange) and the input TOD (blue). The time-streams include the HWPSS at $2f_{\mathrm{HWP}}$. The effect of saturation results in the distortion of the peaks.}
\end{figure}

When the incoming optical power approaches the saturation power, the TES detector begins to saturate and exibit strong nonlinearity, as can be observed in \cref{fig:tes_curve}, where we show the TES response reconstructed according to the procedure described in the companion paper\cite{2024tdh}. The effect of saturation on the TOD can be observed in \cref{fig:tod_comp}, where the TES nonlinear response is simulated following \cref{eq:nl}. Here we see how the cosine wave signal is distorted, as its peaks are slightly flattened. We can also easily predict another effect combining \cref{eq:hwpss} and \cref{eq:nl}. If we consider a generic sinusoidal signal at frequency $\omega$, then it will be up-modulated to $2\omega$ due to TES nonlinearity. It is then clear that nonlinearity in presence of a HWPSS at $2f_{\mathrm{HWP}}$ will produce spurious signals around the science band. We simulate this effect in the time domain, providing an estimation of the required level of knowledge of $g_1$, given the systematic error budget on $r$.

\section{Simulation procedure}
\begin{figure} [ht]
   \begin{center}
   \includegraphics[width=0.5\textwidth]{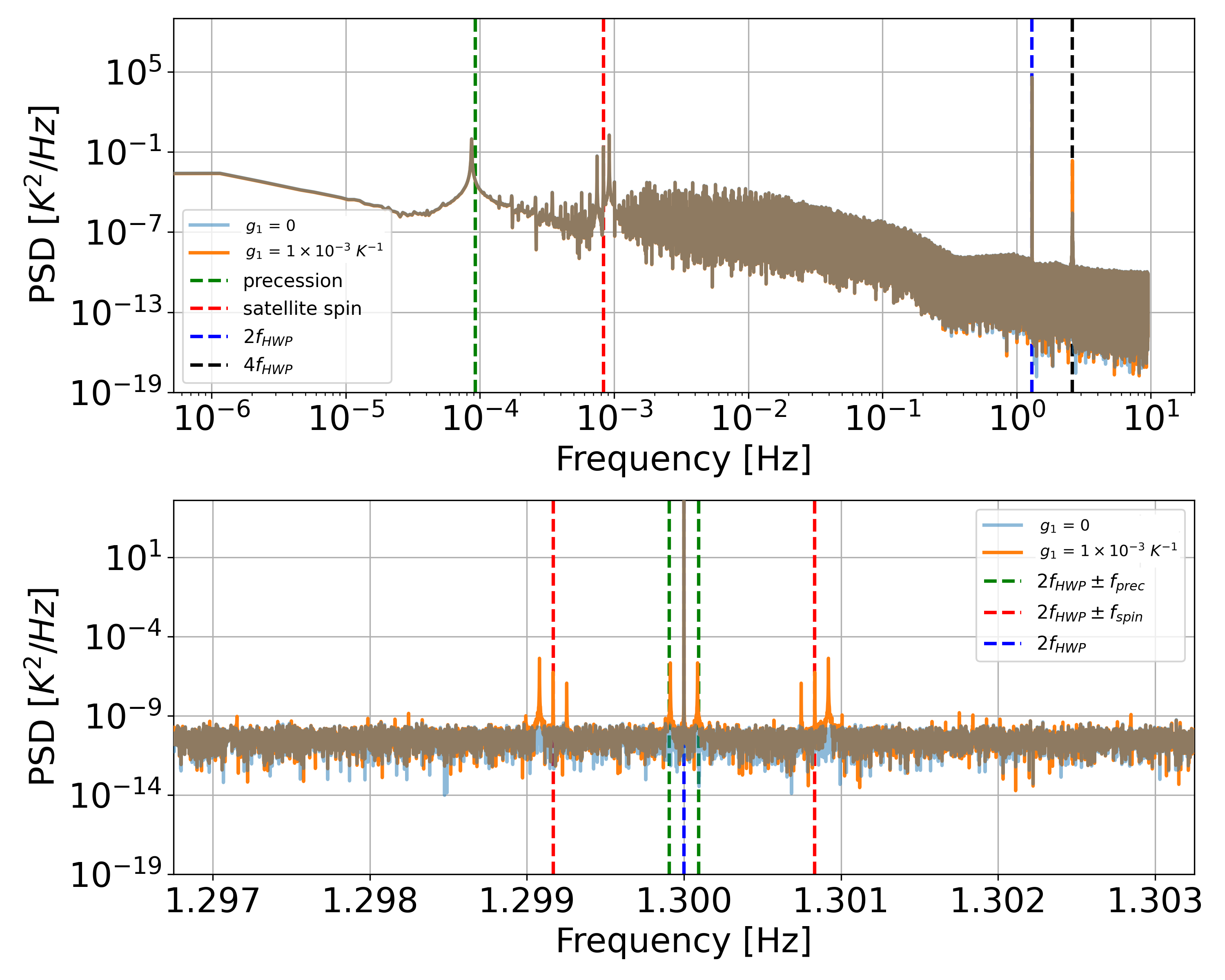}
   \end{center}
   \caption[TOD] 
   {\label{fig:spectra_comp}Effect on the TODs frequency spectra. In orange, with the effect of nonlinearity. The lower panel shows a zoom which highlights the presence of coupling between the dipole and the $2f_{\mathrm{HWP}}$ lines in the frequency domain.}
\end{figure}

We simulate \lb\ one-year observations of the sky, including the signal from the CMB, orbital and solar dipole, synchrotron and thermal dust emission(d0s0 \url{PySM} foregrounds model\cite{2017thorne}), white and 1/$f$ noise and the HWPSS, computed as described in \cref{sec:hwp_diff}. We apply the \lb\ scanning strategy\cite{ptep} for one pair of orthogonal detectors in each observing band. We assume a Gaussian distribution for the nonlinearity parameter, with $\sigma^2_{g_1}$ variance. We inject the nonlinearity in the TOD as described in \cref{sec:tes_nl}, so that the $i$-th realization of the systematic error can be written as:
\begin{equation}
    d_{NL}(t)=[1+{g^i_1} d(t)]d(t)
\end{equation}
If nonlinearity is corrected, for example from ground measurements, then $g_1$ can be interpreted as the level of residual nonlinearity. \\ First of all, it is interesting to analyze the data in frequency space. In presence of nonlinearity, all the periodic components of the signal acquire higher harmonics (\textit{n}=2 only in our second order approximation). This is the case of the acquired $4f_{\mathrm{HWP}}$ peak from HWPSS up-modulations, as can be seen from the upper panel in \cref{fig:spectra_comp}. Besides that, different peaks relative to different periodic signals are coupled as well, as a consequence of nonlinearity. An example of this is the case of the CMB dipole, which appears in the Fourier spectrum at the scanning frequency. We will refer to the frequency of this periodic signal as $f_{\text{dipole}}$. It can be observed in the upper panel of \cref{fig:spectra_comp} as three peaks around $10^{-3}$ Hz. In the lower panel of \cref{fig:spectra_comp}, we see the spikes at $2f_{\mathrm{HWP}} \pm f_{\text{dipole}}$ in the nonlinear signal (in orange).

\subsection{Procedure to set the requirements}
\begin{figure} [ht]
   \begin{center}
   \begin{tabular}{c} 
   \includegraphics[width=0.5\textwidth]{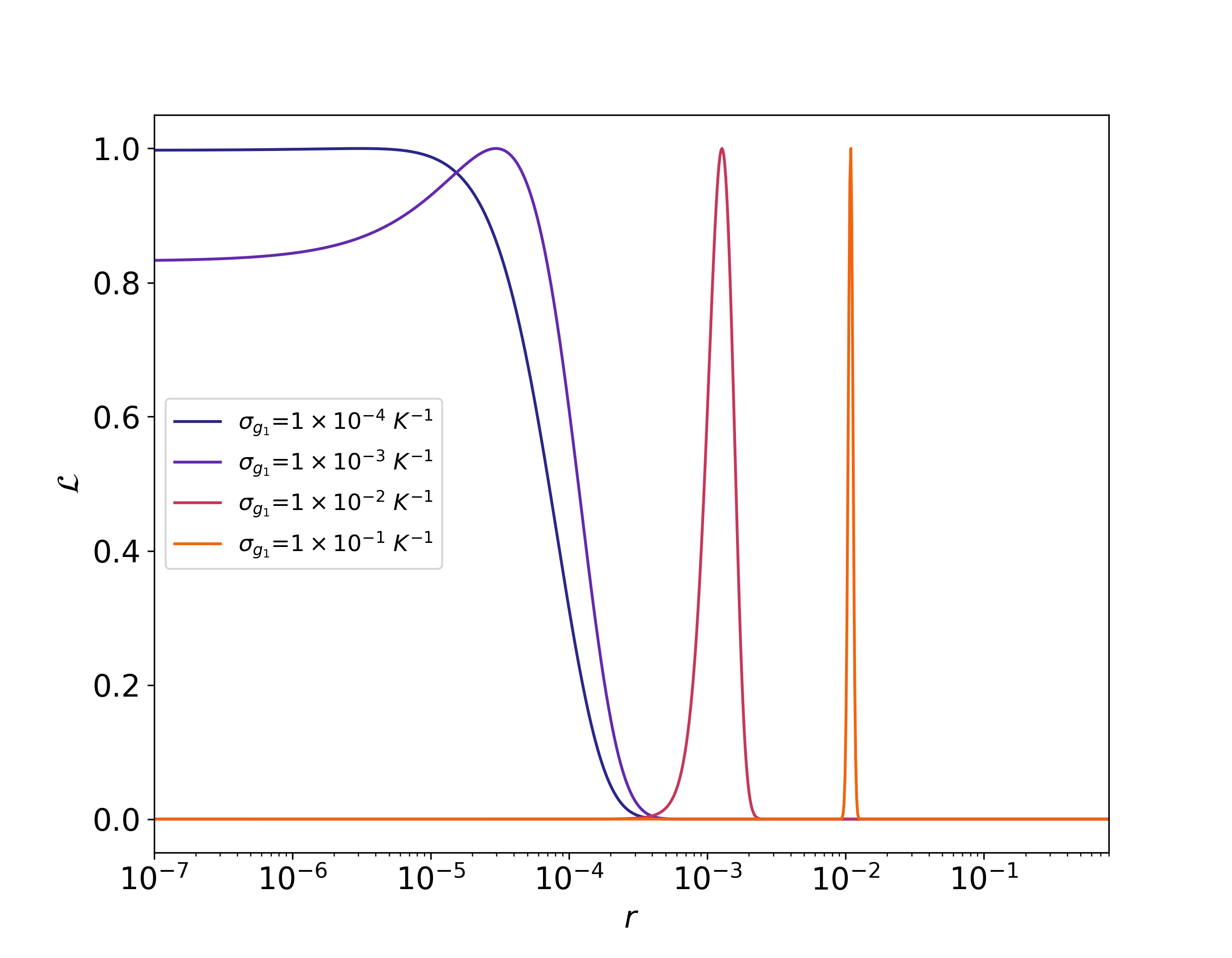}
   \end{tabular}
   \end{center}
   \caption[res] 
   {\label{fig:likes}Example of likelihood of $r$ for different amplitudes of residual nonlinearity studied in this work.}
\end{figure}

We evaluate the impact of residual nonlinearity on the estimation of the tensor-to-scalar ratio with the following pipeline: (i) We produce 100 TODs for each band as described in the previous section. Each uses a different realization of the CMB and noise. The noise has a white and 1/$f$ component (assuming a common knee frequency of 20 mHz\cite{ptep}). (ii) These TODs are simulated both with and without systematics. We assume the same distribution for the uncertainty on the nonlinearity, $\sigma_{g_1}$, for each MHFT channel when producing the contaminated TODs. (iii) We apply a $4f_{\mathrm{HWP}}$ notch filter to evaluate the contribution of nonlinearity in the sidebands. The filter is defined as:
\begin{equation}
    F(f) = 
    \begin{cases}
        1 & f\leq f_1 \\
        0.5(1+\cos{(\pi \frac{f-f_1}{f_2-f_1})}) & f_1<f\leq f_2 \\
        0 & f_2<f\leq f_3 \\
        0.5(1-\cos{(\pi \frac{f-f_3}{f_4-f_3})}) & f_3<f\leq f_4 \\
        1 & f > f_4
    \end{cases}
\end{equation}
where $f_1 = 4f_{\mathrm{HWP}} - 2\delta f$, $f_2 = 4f_{\mathrm{HWP}} - \delta f$, $f_3 = 4f_{\mathrm{HWP}} + \delta f$ and $f_4 = 4f_{\mathrm{HWP}} + 2\delta f$. The amplitude of the filter is set to $\delta f = 4f_{\mathrm{HWP}}/100$. (iv) We produce binned maps from our TODs and we perform a minimum variance component separation to recover CMB maps. (v) We consider the residuals between the recovered maps and the input CMB. \\Details of the simulations are provided below:
\begin{itemize}
    \item We set the distribution of the nonlinearity to be normal with variance $\sigma^2_{g_1}$. We select four values for $\sigma_{g_1}$, ranging logarithmically from $10^{-4}$ to $10^{-1}$ K$^{-1}$. We inject the systematics in all MHFT channels simultaneously. LFT channels are assumed to be ideal. 

    \item We use the HILC (Harmonic Internal Linear Combination\cite{Tegmark2003}) algorithm in \url{FGBuster}\cite{2009stompor} to perform component separation. An analysis using parametric component separation methods resulted in a more stringent requirement, due to strong sensitivity to gain knowledge. We highlight that this analysis is based on the assumption of uniform foregrounds spectral energy density for each component, which is useful to separate the effects of systematics from the complexity of the foreground emission. This assumption could be relaxed, knowing that the foreground spectral parameters vary across the sky\cite{2014fuskeland}. In this case, more complex component separation techniques will be needed to recover CMB maps accurately.

    \item After component separation, we recover the CMB maps, $m^{rec}_{\mathrm{CMB}}$. After subtraction of the input CMB maps, $m^{in}_{\mathrm{CMB}}$, the recovered maps will contain the residual noise, foregrounds and, if present, the effect of systematics. 

    \item We compute the \textit{B}-mode angular power spectrum for $m^{rec}_{\mathrm{CMB}}-m^{in}_{\mathrm{CMB}}$, for both ideal and nonlinear maps. 

    \item We recover the tensor-to-scalar ratio value from the likelihood\cite{Hamimeche2008}: \begin{equation}\label{eq:likel}
    -2\mathcal{L}(C^{BB}_{\ell,\mathrm{obs}}|r) = (2\ell +1)f_{sky}\bigg[\ln{(C^{BB}_{\ell,\mathrm{mod}})} + \frac{C^{BB}_{\ell,\mathrm{obs}}}{C^{BB}_{\ell,\mathrm{mod}}} - \frac{2\ell-1}{2\ell+1}\ln{(C^{BB}_{\ell,\mathrm{obs}})}\bigg]
    \end{equation}

    where 
    \begin{equation*}
        C^{BB}_{\ell,\mathrm{obs}} = C^{\mathrm{res}}_{\ell} + C^{\mathrm{lens}}_{\ell}
    \end{equation*}
    contains the residuals \textit{BB} power spectrum and the lensing \textit{B}-mode spectrum, while
    \begin{equation*}
        C^{BB}_{\ell,\mathrm{mod}} = rC^{\mathrm{GW,}r=1}_{\ell} + C^{\mathrm{lens}}_{\ell} + C^{{\mathrm{{FGs,res}}}}_{\ell} + N_{\ell}
    \end{equation*}
    also accounts for the primordial \textit{B}-mode spectrum for $r=1$, the noise spectrum and the residuals foregrounds after component separation. Examples of the likelihoods in the presence of nonlinearity are shown in \cref{fig:likes}. We extract the 68\%-$\mathcal{L}$ value of $r$, for each realization both with and without systematics. We will refer to the two values as $r_{sys}$ and $r_{nosys}$, respectively. We define the bias on $r$ as $\delta r = |r_{sys}-r_{nosys}|$.

    \item We extract the value of the bias for each $\sigma_{g_1}$, considering the average of $\delta r$ over 100 realizations, $\bar{\delta r}$.

    \item We recover the relation between $\sigma_{g_1}$ and $\bar{\delta r}$ and set the requirement, i.e. an upper limit for the amplitude of the systematic, given the allocated systematic error budget, $6.5 \times 10^{-6}$.
\end{itemize}

\section{Results}
\begin{figure} [ht]
   \begin{center}
   \begin{tabular}{c} 
   \includegraphics[width=0.7\textwidth]{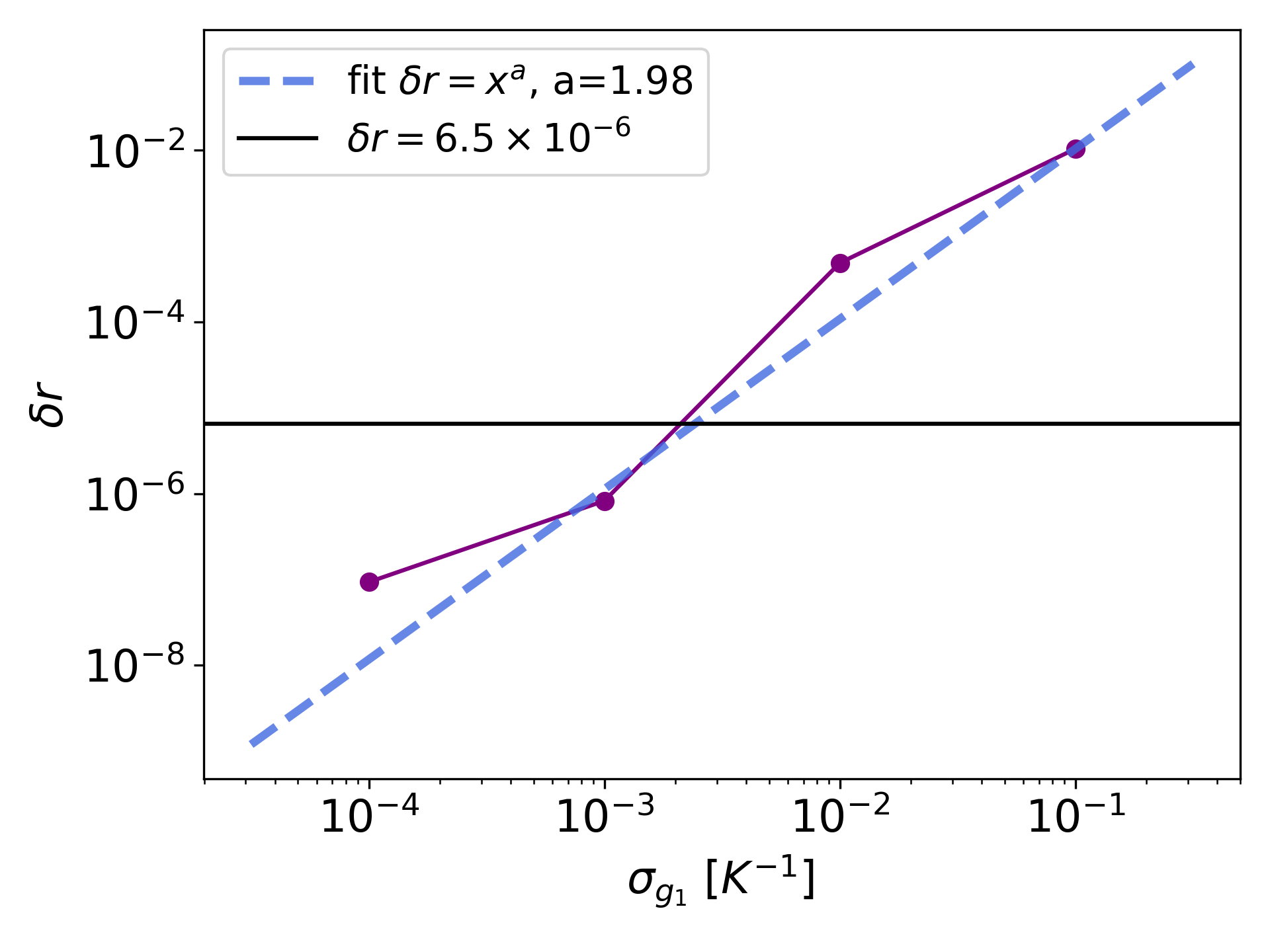}
   \end{tabular}
   \end{center}
   \caption[res] 
   {\label{fig:results}Average $\delta r$ over 100 simulations for different values of nonlinearity. The black solid line marks the threshold of the allocated systematic error budget. The blue dashed line shows the best fit, $\delta r \propto \sigma^2_{g_1}$.}
\end{figure}

Following the procedure described above, we derive a preliminary requirement on the amplitude of the systematics. For $\delta r = 6.5 \times 10^{-6}$, corresponding to the allocated error budget for systematics for \lb\cite{ptep}, we find the requirement on $\sigma_{g_1}$, which should be less than $2 \times 10^{-3}$ K$^{-1}$ as shown in \cref{fig:results}. We also show a quadratic fit as the blue dashed line, verifying $\delta r \propto \sigma^2_{g_1}$, as expected, since $\delta r \propto C_{\ell} \propto \sigma^2_{g_1}$. These results are obtained from simulations of observations of one pair of orthogonal detectors for each observing band. Simulations from single detector observations provide way more stringent requirements. This is due to the fact that when observing the orbital dipole with one detector only, we are not able to correctly resolve the polarization. This results in an intensity-to-polarization leakage of the order of 1 $\mu$K on polarization maps. On the other hand, we expect full focal plane simulations to provide a more relaxed requirement, since we expect the total effect to average down when considering a higher number of detectors. Other mitigation techniques that could be applied to relax the requirement consist of map-based template fitting or broader filtering of the signal about 4$f_{\mathrm{HWP}}$. A pipeline including these options is under development and will be discussed in a future paper. Finally, we point out that these requirements are related to the amplitudes of the HWPSS, which depend on the chosen design for the HWP for MHFT. Thus, the requirements derived in this work could change depending on the value of $A_{2f}$. A combined analysis that lets the amplitude parameter vary as well is now being carried out.

\section{Conclusions}
In this work, we study how the combination of HWP and TES nonidealities will impact the performance of \lb\ in the detection of the tensor-to-scalar ratio. We focus on the Mid- and High- Frequency Telescopes features, considering their different HWP optical properties along the inductive and capacitive axes. We show that the major consequence of this nonideality is the production of a $2f_{\mathrm{HWP}}$ modulation of the sky signal, mainly driven by differential emission. We show how the combination of the HWPSS with TES nonlinearity causes the up-modulation of the $2f_{\mathrm{HWP}}$ to the $4f_{\mathrm{HWP}}$ science band. Moreover, we observe how nonlinearity leads to coupling between periodic signals as well, including the CMB dipole at the scanning frequency. With this work, we assess the needed level of control on nonlinearity in order to meet \lb's systematic error budget. We perform full TOD analyses, injecting the systematics as a second order correction to the TES responsivity. We simulate several multi-frequency observations for different amplitudes of the systematics. CMB residual maps are obtained after component separation and are used to extract the CMB spectra. The \textit{B}-mode spectra are then plug into the $r$ likelihood to extract the induced bias on the tensor-to-scalar ratio, $\delta r$. By imposing $\delta r \leq 6.5 \times 10^{-6}$, we find a preliminary requirement on the level of knowledge of nonlinearity.\\ This work paves the way for further investigations, for example involving full focal plane simulations and including LFT HWP nonidealities. These results are dependent on the amplitude of the $2f_{\mathrm{HWP}}$ signal, that is on the chosen HWP design. A combined analysis, where the HWPSS amplitude is set as an additional free parameter, is under development.

\acknowledgments 

%
We acknowledge the use of \url{litebird_sim} software package and the use of computing resources at CINECA.
\textit{LiteBIRD} (phase A) activities are supported by the following funding sources: ISAS/JAXA, MEXT, JSPS, KEK (Japan); CSA (Canada); CNES, CNRS, CEA (France);
DFG (Germany); ASI-Grants No. 2020-9-HH.0 and 2016-24-H.1-2018, INFN, INAF (Italy); RCN (Norway); MCIN/AEI, CDTI (Spain); SNSA, SRC (Sweden); and NASA, DOE (USA).
%

\bibliography{report} 
\bibliographystyle{spiejour} 

\end{document}